\documentstyle[12pt,epsf]{article}
\catcode`@=11
\def\@cite#1{$^{\scriptsize{#1}}$}
\def\@refe#1{#1}
\def\@biblabel#1{{\normalsize\bf{#1}}}
\def\refe{\@ifnextchar
[{\@tempswatrue\@citexr}{\@tempswafalse\@citexr[]}}
\def\@citexr[#1]#2{\if@filesw\immediate\write\@auxout{\string\citation
{#2}}\fi
  \def\@citea{}\@refe{\@for\@citeb:=#2\do
    {\@citea\def\@citea{,}\@ifundefined
       {b@\@citeb}{{\bf ?}\@warning
       {Citation `\@citeb' on page \thepage \space undefined}}%
\hbox{\csname b@\@citeb\endcsname}}}{#1}}
 \catcode`@=12
\title{$12j$-symbols and four-dimensional \\quantum gravity }
\author{Mauro Carfora\dag \S , Maurizio Martellini\ddag \S ,\\
 Annalisa Marzuoli\dag \S\\
\dag Dipartimento di Fisica Nucleare e Teorica dell'Universit\`a di
Pavia\\
Via Bassi 6, I-27100 Pavia, Italy \\
\ddag
Dipartimento di Fisica, Universit\`a di Roma I {\it La Sapienza} \\
Piazzale Aldo Moro 2, 00185 Roma, Italy  \\
On leave of absence from: \\
Dipartimento di Fisica dell'Universit\'a di Milano,\\
Via Celoria, 16, I-20133 Milano, Italy \\
\S Istituto Nazionale di Fisica Nucleare, Sezione di Pavia, Italy\\}
\date{}
\begin{document}
\maketitle
\begin{abstract}
\noindent We propose a model which represents a four-dimensional
version of Ponzano and Regge's three-dimensional euclidean
 quantum gravity. In particular we show that the exponential
of the euclidean Einstein-Regge action for a $4d$-discretized
block is given, in the semiclassical limit, by a
gaussian integral of a suitable $12j$-symbol. Possible
developments of this result are discussed.
\end{abstract}
\vskip7cm
\centerline{Preprint FNT/T 92/37 (October 1992)}
\vfill\eject

In 1968 Ponzano and Regge\cite{PR} discovered a deep connection
between the expansion of  a Racah-Wigner $6j$-symbol for large
values of its angular momenta and the partition function for $3d$-euclidean
quantum gravity, discretized according to Regge's
prescription\cite{R1}. In the early 80's other authors\cite{{HP},{L}}
added some interesting results to the original idea, but it is just
during the last year that new exciting results have been
provided\cite{{TV},{OS},{AW},{MT}}.\par
\noindent Before addressing our main topic, and in order
to fix notations which will be used in the following,
we give a brief review of Ponzano and Regge's model. The asymptotic
form of the Racah-Wigner $6j$-symbol reads\cite{PR}:\par
\begin{eqnarray}
\left\{\begin{array}{ccc}
J_1 & J_2 & J_3 \\
J_4 & J_5 & J_6
\end{array}\right\}
\buildrel {J_m \gg 1} \over
\longrightarrow
\frac{1}{\sqrt{12\pi V_T}}
\cos( S_R[T] + \pi/4 )
\label{1}
\end{eqnarray}
where $J_m=0,1/2,1,3/2,2,\ldots$ for each $m=1,2,\ldots,6$ and units are chosen
for which $\hbar =G=1$ ($G$ is the Newton constant).
 The geometrical interpretation of the right-hand side
of the former expression relies on the tetrahedral symmetry of the
$6j$-symbol. The six quantities $(J_m + 1/2)$, $m=1,2,\ldots,6$, can be
associated with the edges in the surface of a tetrahedron $\tau$ embedded
in ${\bf R}^3$. $T$ is the tetrahedron (more precisely, the 3-simplex)
obtained by filling in the surface $\tau$ with a portion of flat
3-space (see fig. 1). $V_T$ is the euclidean volume of $T$, and
finally $S_R[T]$ is the euclidean Einstein-Regge action for $T$,
namely\cite{R1}:
\begin{eqnarray}
S_R[T]=\sum_{m=1}^{6}{(J_m+1/2)\theta_m}
\label{2}
\end{eqnarray}
where $\theta_m$ is the angle between the outer normals of the
two faces of $T$ which share the $m$-th edge. (Recall that the euclidean
Einstein-Regge action in dimension $d$ is a discretized version\cite{R1}
 of the
usual Einstein-Hilbert  action $\int {d^dx\sqrt{g}R}$ for a smooth
riemannian manifold $({\cal M}^d,g)$ when one takes a simplicial
decomposition $M^d$ into euclidean $d$-simplices; the curvature is
associateted with the collection of the $(d-2)$-simplices, or bones).\par
\noindent In the limit $J_m \gg1$ , we may disregard $1/2$ with respect to
$J_m$ in (2), and thus rewrite (1) (up to a constant phase factor)
simply as:
\begin{eqnarray}
\left\{\begin{array}{ccc}
J_1 & J_2 & J_3 \\
J_4 & J_5 & J_6
\end{array}\right\}
\sim "PFP"\,( exp \, \{i \sum_{m=1}^{6}J_m\theta_m\})
= "PFP"\,( exp \, \{i S_R[T]\})
\label{3}
\end{eqnarray}
where $"PFP"$ means "take the positive frequency part of
(\ldots )", and the symbol $\sim$ stands, from now on, for
the limit $J_m \gg 1$.
\noindent Expression (3) represents the semiclassical limit of the
partition function for $3d$-euclidean Einstein-Regge gravity involving
just one elementary building block, namely the tetrahedron $T$. In order
to consider the case of a generic $3d$-simplicial manifold $M^3$ of
fixed topology, one has to perform the following steps: {\it i})
associate with each $3$-simplex in $M^3$ a $6j$-symbol;
{\it ii}) take the sum over the "internal edges" of the
simplicial dissection of the product of these $6j$'s and {\it iii})
multiply by suitable additional factors. After a regularization,
one gets the semiclassical partition function for $M^3$. This procedure
is extensively illustrated elsewhere\cite{{PR},{HP},{O}},
 so that we do not insist anymore on this point.\par
One of the main open problem connected with the procedure outlined
above is the impossibility of extending it to higher-dimensional
cases, and in particular to the physically significant
case of $4d$-quantum gravity. This paper represents a first progress
in this direction. In particular, we shall provide a $4d$ version of the
asymptotic formula (3) involving in its right-hand side the
euclidean Einstein-Regge action for a suitable $4d$-simplex.
On the left-hand side there will be a gaussian integral of a
$12j$-symbol according to the procedure which we are going to describe.\par
\noindent The $3nj$-symbols $(n=2,3,\ldots)$ appear in the quantum theory
of angular momentum when one performs the decomposition of tensor
representations of the group $SU(2)$ into the direct sum of
irreducible representations. Consider in particular the $12j$-symbol
of the first kind (we adopt from now on the notation of
Yutsis {\it et al.}\cite{Y}):
\begin{eqnarray}
\left\{\begin{array}{cccc}
j_1 & j_2 & j_3 & j_4 \\
l_1 & l_2 & l_3 & l_4 \\
k_1 & k_2 & k_3 & k_4
\end{array}\right\}
\label{4}
\end{eqnarray}
where all $j$'s, $l$'s and $k$'s run over $\{0,1/2,1,3/2,2,\ldots\}$.\par
\noindent Upon imposing the condition $j_4=0$, one gets what we shall
call the {\it reduced} $12j$-symbol:
\begin{eqnarray}
\left\{\begin{array}{cccc}
j_1 & j_2 & j_3 & 0 \\
l_1 & l_2 & l_3 & l_4 \\
k_1 & k_2 & k_3 & k_4
\end{array}\right\}
\equiv
\left\{\begin{array}{c}
reduced \\
12j
\end{array}\right\}
\label{5}
\end{eqnarray}
where the shorthand notation in the right-hand side will be used
whenever the specification of the arguments will be clear from the
context.\par
\noindent The symmetry properties of the reduced $12j$-symbol
tell us that:
\begin{eqnarray}
\left\{\begin{array}{c}
reduced \\
12j
\end{array}\right\}=0 \; \mbox{unless}: l_4=k_1 \, \mbox{and} \: l_3=j_3
\label{6}
\end{eqnarray}
from which we see that only $9$ among the original parameters
are independent.\par
\noindent The reduced $12j$-symbol can be related to a product of two
$6j$-symbols by means of an exact relation\cite{Y} which, after
a suitable change of variables, reads:
\begin{eqnarray}
\left\{\begin{array}{cccc}
J_1 & J_2 & J'_2 & 0 \\
J_3 & J'_3 & J'_2 & J_5 \\
J_5 & J_4 & J'_4 & J_6
\end{array}\right\} = N \,  \delta (J_4,J'_5) \delta (J_2,J'_1)
 \delta (J_6,J'_6)
\left\{\begin{array}{ccc}
J_1 & J_2 & J_3 \\
J_4 & J_5 & J_6
\end{array}\right\}
\left\{\begin{array}{ccc}
J'_1 & J'_2 & J'_3 \\
J'_4 & J'_5 & J'_6
\end{array}\right\}
\label{7}
\end{eqnarray}
where:
\begin{eqnarray}
N = (-1)^{J_5-J_4+J'_3+J'_2-J_2+J_1} /  \sqrt{J_5J'_2}
\label{8}
\end{eqnarray}
Owing to the presence of the three Kronecker delta's in (7), among
the twelve angular momenta $\{J_m, J'_n\},(n,m=1,2,\ldots ,6)$, only
the correct number of independent $J$'s and $J'$'s survives, namely:
\begin{eqnarray}
\{J_1,J_2,J_3,J_4,J_5,J_6,J'_2,J'_3,J'_4\}
\label{9}
\end{eqnarray}
The form of the decomposition (7) allows to evaluate the
semiclassical limit of the reduced $12j$-symbol simply by applying
the result of Ponzano and Regge (3) to each $6j$-symbol. Then
we have:
\begin{eqnarray}
\left\{\begin{array}{c}
reduced \\
12j
\end{array}\right\} \sim
\frac{N \, \delta (a;b;c)}{\sqrt {12\pi V_T} \cdot \sqrt {12\pi V_{T'}}}
\, exp \left\{ i\sum_{m=1}^{6}J_m \theta_m +i \sum_{n=1}^{6}J'_n \theta '_n
\right\}
\label{10}
\end{eqnarray}
where: $\sim$ stands for $J_m$, $J'_n \gg 1$, $(m,n=1,2,\ldots ,6)$;
$a\equiv J_4-J'_5$, $b\equiv J_2 - J'_1$, $c\equiv J_6-J'_6$;
$\delta (a;b;c)=1$ if and only if $\delta (a)=\delta (b)=\delta (c)=
1$, and $\delta (a;b;c)=0$ otherwise. $V_T[V_{T'}]$ is the volume
of the tetrahedron $T[T']$ associated with the $6j$-symbol
containing $\{ J_m \}[\{ J'_n\}]$; $\theta_m [\theta '_n]$ is the
angle between the outer normals of the
two faces of $T[T']$ sharing the edge $J_m[J'_n]$.\par
\noindent Notice that in (10) we have omitted "$PFP$ of the
exponential" which appeared in (3); as we shall see, this is completely
consistent within the framework of our procedure. Moreover, the
argument of the exponential can be rearranged in order to take
into account (9). To this end, we first introduce a relabelling
of the nine angular momenta, namely:
\begin{eqnarray}
\left( J_1, J_2, J_3, J_4, J_5, J_6, J'_2, J'_3, J'_4 \right) \rightarrow
\left( {\cal J}_1, {\cal J}_2, {\cal J}_3, \ldots , {\cal J}_9 \right)
\label{11}
\end{eqnarray}
Then, according with this new enumeration, we can define:
\begin{eqnarray}
\left( \theta_1 , (\theta_2 +\theta '_1),\theta_3 , (\theta_4 + \theta '_5),
\theta_5 , (\theta_6 +\theta '_6), \theta '_2, \theta '_3, \theta '_4 \right)
\equiv \left( \psi_1 , \psi_2 , \psi_3 , \ldots , \psi_9 \right )
\label{12}
\end{eqnarray}
With these substitutions (10) becomes:
\begin{eqnarray}
\left\{\begin{array}{c}
reduced \\
12j
\end{array}\right\} \sim \tilde{N} \;exp \left\{ i \sum_{k=1}^{9}
{\cal J}_k \psi_k \right\}
\label{13}
\end{eqnarray}
where we took into account the fact that $\delta (a;b;c)=1$ and
we set $\tilde{N} \equiv N/\sqrt{12\pi V_T} \cdot \sqrt{12\pi V_{T'}}$.\par
\noindent For a given real, non singular and symmetric $9 \times 9$ matrix
$\Delta$ consider now the following multiple gaussian integral over
the set of real variables $\{\psi_i\}$, $(i=1,2,\ldots ,9)$:
\begin{eqnarray}
Z\left[ {\cal J}_i \, ;\Delta _{ki}\right] =
{\int
\left\{\begin{array}{c}
reduced \\
12j
\end{array}\right\}
exp \: \left\{- \frac{1}{2} \, \sum_{k,i=1}^{9} \psi _k \Delta _{ki} \,
\psi _i\right\} \prod_{j=1}^{9} d\psi _j}
\label{14}
\end{eqnarray}
Replacing (13) in the above expression, and using the standard formula
for gaussian integration, we find that in the semiclassical limit
(i.e. in the limit ${\cal J}_i \gg 1$)the following result holds
true:
\begin{eqnarray}
Z\left[ {\cal J}_i \, ;\Delta _{ki}\right] \sim \hat{N}
\left( det\Delta \right)^{-1/2} exp \: \left\{  -\sum_{k,i=1}^{9}
{\cal J}_k (\Delta ^{-1})_{ki} \, {\cal J}_i \right\}
\label{15}
\end{eqnarray}
where $det\Delta$ and $\Delta ^{-1}$ are respectively
the determinant and the inverse of the matrix $\Delta$,
and we put $\hat{N} \equiv (2\pi )^6 \tilde{N}$.
\par
\noindent In order to interpret (15), recall that the form of the
euclidean Einstein-Regge action for a generic $4d$-simplicial
manifold $M^4$ with boundary $\partial M^4$ is (up to an arbitrary
term depending on the edge lenghts in the boundary)\cite{HS}:
\begin{eqnarray}
8\pi S_R[M^4]=\sum_{b\in int M^4} A(b)\epsilon (b) +
\sum_{b\in \partial M^4} A(b) \alpha (b)
\label{16}
\end{eqnarray}
where $b$ stands for "bone" ( in dimension $4$ a bone is a $2$-simplex
where, according to Regge Calculus, curvature is concentrated) and
$A(b)$ is the area of $b$. The first sum in (16) is over the collection
of bones belonging to the interior of $M^4$, $int M^4$, and
$\epsilon (b)$ represent the so called defect angle\cite{R1}
 associated with
the bone $b$. The second sum is over the bones lying in
the boundary of $M^4$, $\partial M^4$, and $\alpha (b)$
has to be interpreted as the angle between the outer normals
of the two boundary $3$-simplices which intersect at $b$.\par
\noindent Consider again our expression (15), and in particular
the sum appearing as the argument of the exponential.
The terms of this sum are
quadratic in the angular momenta ${\cal J}_i$, $(i=1,2, \ldots ,9)$,
and thus are related to the area of some $2$-dimensional
geometric object. This circumstance relies on the geometrical interpretation
of the reduced $12j$-symbol.
Begin then by recalling that, just like in the
case of a $6j$-symbol, the structure of the reduced $12j$-symbol
(5) can be associated with a diagram in ${\bf R}^3$ (indeed,
this is true for any $3nj$-symbol\cite{{PR},{Y}}).
As is easily seen from fig.2a, this diagram is given by the pair
of tetrahedra $T$ and $T'$ joined along one of their face
({\it cfr.} also decomposition (7)).\par
\noindent The effect on the
reduced $12j$-symbol of the gaussian integration performed
in (14), at the semiclassical level (15), is to give rise to terms
proportional to the product $J_m \cdot J'_n$, for suitable
$m$ and $n$ ({\it cfr.} (11)).
The crucial observation is that the existence of such contributions
can be explained in a coherent way only by allowing the appearance
of a tenth edge besides the nine original ${\cal J}_i$. The role
of this new edge, the lenght of which will be denoted by $L$,
becomes more transparent if we refer to fig.2b, where we show
a representation  of an euclidean $4$-simplex
$\sigma$ which is built up from the $3d$-simplicial manifold
of fig.2a by joining in ${\bf R}^4$ the upper and the lower vertices
with an edge of lenght $L$. Then we see that $\sigma$ has three
faces, namely $(J_1 L J'_2)$, $(J_3 L J'_3)$, $(J_5 L J'_4)$,
which contain the edge L and which appear in (15) in three terms,
proportional to $J_1 \cdot J'_2$, $J_3 \cdot J'_3$ and
$J_5 \cdot J'_4$ respectively.\par
\noindent  Then the sum in the exponential (15) can be related to the
Einstein-Regge action (16) written for the $4$-simplex $\sigma$
provided that we identify (up to some normalization factors, see
below) the matrix elements $(\Delta ^{-1} )_{ki}$ with a suitable set of
angles of the type  $\epsilon (b)$ and $\alpha (b)$. More
precisely, we set:
\begin{eqnarray}
\sum_{k,i=1}^{9} {\cal J}_k(\Delta ^{-1})_{ki}\,{\cal J}_i \equiv
\sum_{k,i=1}^{9} \Theta _{ki} A_{ki}
\label{17}
\end{eqnarray}
where $A_{ki}$ is the area of the $2$-face of $\sigma$ containing
${\cal J}_k$ and ${\cal J}_i$ (up to
combinatorial factors). According to (16), the angle $\Theta _{ki}$
has to be interpreted either as a defect angle (if the $2$-face
containing ${\cal J}_k$ and ${\cal J}_i$
belongs to the interior of $\sigma$), or as an angle between the
outer normals of the two $3$-simplices which share the
$({\cal J}_i ,{\cal J}_k)$-face (if this $2$-face belongs
to the boundary of $\sigma$). For what concerns the explicit
relation between the matrices $\Delta ^{-1}$ and $\Theta$, notice
that the area $A_{ki}$ is, for each $k,i=1,2,\ldots ,9$, a known
function of the edges\cite{H} and also that obviously $A_{ii}=0$
( the reader may find elsewhere\cite{RW} more information
about the properties of the matrix $\Theta$ ).\par
\noindent Then, according to (15) and (17) we can write :
\begin{eqnarray}
Z[{\cal J}_i\,;\Delta _{ki}] \, \sim \, \tilde{N} (det\Delta)^{-1/2}\;
exp\, \{-S_R[\sigma]\}
\label{18}
\end{eqnarray}
where $S_R[\sigma]$ is the Einstein-Regge action for the $4$-simplex
$\sigma$.\par
\noindent As we anticipated before, the right-hand side of
this expression represents the semiclassical limit of the partition
function for $4d$-gravity and can be considered as a direct
 generalization of Ponzano and Regge's result (3).
There is however an important difference. In our expression the
euclidean action $S_R[\sigma]$ appears in the exponential
multiplyied by the exact factor needed in order to interpret
the right-hand side of (18) as the semiclassical limit of an
euclidean partition function. That is quite remarkable, as either
in Ponzano and Regge's or in Turaev and Viro's models the
presence of the imaginary unit $i$ in front of the euclidean
action has not been yet explained in a completely satisfactory
way.\par
\vskip0.5cm
Let us turn now to possible developments connected to our
result (18). An important further step would be to show that
a scaling relation similar to the existing relation for Ponzano
and Regge's model (based on the Biedenharn-Elliot identity)
 holds true\cite{PR}. Moreover, our result represents just the
"elementary building block" for a $12j$-model of $4d$-euclidean
gravity, so that another point to deal with should concern the
possibility of writing a partition function for a generic
$4d$-combinatorial manifold dissected into $4$-simplices.\par
\noindent Obviously, the most important issue to address
concerns the set up of a regularization procedure yielding
for a consistent continuum limit of our model. In the original
$3d$-case of Ponzano and Regge such step has been implemented
by exploiting the Turaev and Viro's quantum
 $6j$-model\cite{{TV},{OS},{AW},{MT}}.
In this way it has been shown that the semiclassical
continuum limit of the partition function of such q-model defines
a naturally regularized path integral for $3d$-euclidean gravity.
It is not clear if there is a suitable generalization of these models
to our case, mainly because very little ( if not nothing ) appears
to be currently available on quantum $12j$-symbols.\par
\noindent We are addressing this issue, but we have not
 yet been able to provide a definite answer to such questions.
Nonetheless we are confident that this line of attack deeply
probes into the structure of  $4$-d euclidean quantum gravity.
\vskip3cm
{\bf Acknowledgement}\par
We would like to thank Tullio Regge for useful conversations.\par
\vfill\eject

\section*{References}
\begin{description}
\bibitem[1] {PR}
G.Ponzano and T.Regge, {\it Semiclassical limit of Racah coefficients},
in: Spectroscopic and Group Theoretical Methods in Physics, ed. F. Block
{\it et al.} ( North Holland, Amsterdam, 1968) pp. 1-58.

\bibitem[2] {R1}
T.Regge, Nuovo Cimento {\bf 19} (1961) 558-571.

\bibitem[3] {HP}
B. Hasslacher and M.J. Perry, Phys. Lett. {\bf B 103} (1981) 21-24.

\bibitem[4] {L}
S.M. Lewis, Phys. Lett. {\bf B 122} (1983) 265-267.

\bibitem[5] {TV}
V.G. Turaev and O.Y. Viro, {\it State sum invariants of 3-manifolds and
quantum 6j-symbols}, LOMI Preprint (1990).

\bibitem[6] {OS}
H. Ooguri and N. Sasakura, Mod. Phys. Lett. {\bf A 6} (1991)
3591-3600.

\bibitem[7] {AW}
F. Archer and R.M. Williams, Phys. Lett. {\bf B 273} (1991) 438-444.

\bibitem[8] {MT}
S. Mizoguchi and T.Tada, Phys. Rev. Lett. {\bf 68} (1992) 1795-1798.

\bibitem[9] {O}
H. Ooguri, Nucl. Phys. {\bf B 382} (1992) 276-303.

\bibitem[10] {Y}
A.P. Yutsis, I.B. Levinson and V.V. Vanagas, {\it Mathematical
apparatus of the theory of angular momentum} (Israel program
for scientific translations, Jerusalem, 1962).

\bibitem[11] {HS}
J.B. Hartle and R. Sorkin, Gen. Rel. Grav. {\bf 13} (1981) 541-549.

\bibitem[12] {H}
H.W. Hamber, in Proc. of the Les Houches Summer School 1984,
ed. K. Osterwalder and R. Stora ( North Holland, Amsterdam, 1986).

\bibitem[13] {RW}
M. Ro\u{c}ek and R.M. Williams, Phys. Lett. {\bf 104 B} (1981) 31-37.
\end{description}
\vfill\eject

{\bf Figure captions}
\vskip2cm
\noindent {\bf fig. 1}\par
The diagram of the $6j$-symbol is the $3$-simplex $T$ with
boundary $\partial T \equiv \tau$ embedded in ${\bf R}^3$.
 $( T,\tau )$ is homeomorphic to $(D^3, S^2)$, where $D^3$
is the euclidean $3$-disk and $S^2$ is the $2$-sphere.
\vskip1cm
\noindent {\bf fig. 2a}\par
The diagram of the reduced $12j$-symbol is the $3d$-combinatorial
manifold obtained by joining the two tetrahedra $T$ and $T'$.
 Notice that the three edges of $T$ $(J_4, J_2, J_6)$
form a face which is glued to the face $(J'_5, J'_1, J'_6)$ of
$T'$.
\vskip1cm
\noindent {\bf fig. 2b}\par
The $4$-simplex $\sigma$ can be represented in ${\bf R}^3$ as the
combinatorial manifold of fig.2a with an additional edge of
lenght $L$ connecting two vertices as indicated ( the heavy
line $L$ should be thought as lying in the fourth dimension).
 This drawing gives however the correct vertex-edge-face
scheme of $\sigma$ ($\sigma$ has 5 vertices, 10 edges and 10
triangular faces). $\sigma$ is topologically the $4$-disk in
${\bf R}^4$ with boundary the $3$-sphere.

\end{document}